\documentclass[preprint2]{aastex}

\slugcomment{Resubmitted to PASP, October 2005.}

\shorttitle{Rate of Period Change as a Diagnostic of Cepheid Properties}
\shortauthors{Turner, Abdel-Sabour \& Berdnikov}

\begin{document}

\title{Rate of Period Change as a Diagnostic of Cepheid Properties}

\author{David G. Turner}
\affil{Department of Astronomy and Physics, Saint Mary's University,
    Halifax, Nova Scotia B3H 3C3, Canada}
\email{turner@crux.smu.ca}

\author{Mohamed Abdel-Sabour Abdel-Latif}
\affil{Department of Astronomy and Astrophysics, National Research 
    Institute Of Astronomy and Geophysics (NRIAG), Box 11242, 
    Helwan, Cairo, Egypt}
\email{sabour2000@hotmail.com}

\and

\author{Leonid N. Berdnikov}
\affil{Sternberg Astronomical Institute and Isaac Newton Institute of 
    Chile, Moscow Branch, 13 Universitetskij prosp., Moscow 119899, Russia}
\email{berdnik@sai.msu.ru}

\begin{abstract}
Rate of period change $\dot{P}$ for a Cepheid is shown to be a parameter 
that is capable of indicating the instability strip crossing mode for 
individual objects, and, in conjunction with light amplitude, likely 
location within the instability strip. Observed rates of period change in 
over 200 Milky Way Cepheids are demonstrated to be in general agreement 
with predictions from stellar evolutionary models, although the sample 
also displays features that are inconsistent with some published models 
and indicative of the importance of additional factors not fully incorporated 
in models to date.
\end{abstract}


\section{Introduction}

Cepheids represent a brief phase in the post-main-sequence evolution 
of stars that originally had masses in excess of $\sim 3 \frac{1}{2} 
M_{\sun}$ \citep{tu96}. As evolved objects they populate the Cepheid 
instability strip in the HR diagram according to the manner in which 
they generate energy, depending upon strip crossing. Intermediate mass 
stars can become unstable to radial pulsation during shell hydrogen 
burning (first crossing), twice during core helium burning (second and 
third crossings), and according to some evolutionary models 
\citep{ib65,be77,be85,xl04}, twice during shell helium burning (fourth 
and fifth crossings). A common feature of many recent evolutionary 
models for intermediate-mass stars, which incorporate the new opacity 
tables \citep[e.g.,][]{mm00,mm02,bo00,sa00}, is that they permit only 
three strip crossings for such stars, since core oxygen ignition occurs 
prior to a separate shell helium-burning phase.

In all cases the evolution of stars through the Cepheid instability 
strip should be associated with gradual changes in overall dimensions, 
and hence periods of pulsation: period increases for evolution towards 
the cool edge of the instability strip as the stellar radius grows, and 
period decreases for evolution towards the hot edge as the stellar 
radius decreases. The observed parabolic trends in Cepheid O--C diagrams 
(plots of the differences between Observed times of light maximum and 
those Computed from a linear ephemeris) have been recognized for the 
past half century as evidence for the evolution of such stars through 
the instability strip \citep{pa58,st59,ei82}. As noted by \citet{st59}, 
``It appears that studies of period change are by far the most sensitive 
test available to the astronomer for detecting minute alterations in the 
physical characteristics of a star.''

Observations of period changes in Cepheids have been matched with some 
confidence to evolutionary models of massive stars in various crossings 
of the instability strip \citep[e.g.,][]{tu98,tb01,tb04} in order to 
identify the direction of strip crossing for individual variables. When 
used for such purposes, the study of Cepheid period changes becomes an 
important tool for the characterization of individual members of the 
class.

In principle it should also be possible to use rate of period change for 
individual Cepheids to establish likely location within the instability 
strip. Because strip crossings for individual Cepheids occur at different 
rates and at different luminosities for specific stellar masses, the 
observed rates of period change must be closely related to strip crossing 
mode and location within the instability strip. Potential constraints 
are imposed by variations in chemical composition and pulsation mode, 
e.g., fundamental mode, first overtone, etc. \citep{be97,te99}, as well as 
by our limited ability to establish small rates of period change for O--C 
data containing sizeable observational uncertainties \citep{sz83}. In 
this paper we demonstrate the link in more detail.

\section{Basis of the Relationship}

The link between rate of period change in Cepheids and location 
within the instability strip is illustrated with the aid of Fig. 1. 
The diagram is a theoretical HR diagram that depicts the location 
of the Cepheid instability strip according to the parameters derived 
for Milky Way Cepheids \citep{tu01}, along with Geneva evolutionary 
tracks for stars of 4, 5, 7, and 10 $M_{\sun}$ at $Z = 0.008$ from 
\citet{ls01}. Lines of constant stellar radius are shown crossing 
various portions of the instability strip. According to the well 
established Cepheid period-radius relation, they should represent 
lines of constant pulsation period for individual Cepheids.

>From an examination of Fig. 1 it is clear that, if one considers only 
Cepheids of a specific period and in a common crossing of the instability 
strip, those on the hot edge of the strip must be $\sim 20\%$ more 
massive than those on the cool edge of the strip. Since rate of evolution 
increases in proportion to the mass of a star, Cepheids lying on the hot 
edge of the strip are evolving faster, and hence changing their pulsation 
periods at a more rapid rate, than Cepheids of identical period lying on 
the cool edge of the strip. Rate of period change therefore relates 
directly to location within the instability strip for individual Cepheids. 
Differences in strip crossing modes are only a minor concern. Cepheids 
with increasing periods must be in the first, third, or fifth crossing of 
the strip, whereas Cepheids with decreasing periods must be in the second 
or fourth crossing of the strip.

A minor complication arises from restrictions on our ability to identify 
period changes in Cepheids tied solely to stellar evolution. Some Cepheids 
exhibit erratic period changes that appear to originate from random 
fluctuations in pulsation period. SZ Tau \citep{bp95}, S Vul \citep{be94}, 
and V1496 Aql \citep{be04} are excellent examples, although in the first 
two cases it is possible to identify the underlying evolutionary 
modifications to pulsation period.

A study by \citet{bi00} may give the impression that stellar evolution has 
only a minor effect on Cepheid O--C diagrams, since it notes that parabolic 
trends were detected in only 67 of 230 Cepheids surveyed. That number is 
misleading, however, given that a previous survey by \citet{tu98} had 
found parabolic trends in 137 Cepheids from a much smaller sample. It was 
actually intended to indicate the poor temporal coverage and lack of 
extensive O--C data available for many well-studied Galactic Cepheids, a 
situation that has been remedied in recent years by our ongoing program 
to obtain archival data on Cepheid brightness variations using the Harvard 
College Observatory Photographic Plate Collection. At present the parabolic 
trends in O--C diagrams typical of stellar evolution are found to be 
extremely common. A survey by \citet{gl05} cites a typical frequency of 
$\sim 80\%$ in both cluster and field Cepheids, for example, although their 
``anomalous'' objects include Cepheids like SV Vul in which the evolutionary 
trend is quite distinct \citep{tb04}. A more realistic frequency for Milky 
Way Cepheids displaying evolutionary trends is in excess of $\sim 90\%$. For 
many of the remaining objects, the evolutionary trends may be more obvious 
in longer time baselines of light curve coverage.

As also pointed out by \citet{fe90} and by \citet{bt04}, the O--C trends 
indicative of evolution in Cepheids need not be strictly parabolic. If 
the rate at which a massive star is evolving through the instability 
strip is not constant with time, the O--C data for the associated Cepheid 
variable may be better described by a third or fourth order polynomial. 
The Cepheids Y Oph \citep{fe90} and WZ Car \citep{bt04} are two objects 
(of several hundred) where that appears to be the case. Such complications 
may affect the derived rates of period change, but in most cases only by 
small amounts. In the large majority of studies of Cepheid period changes, 
the derived rate of period change reflects the evolution of the star 
through the instability strip \citep[see][]{sz83}.

\section{Stellar Evolution Predictions}

Most computational evolutionary models for evolved stars are used for 
constructing evolutionary tracks rather than testing for pulsation 
instability. But, as noted by citet{pa58}, it is possible to use the basic 
information they provide on gradual changes in luminosity and effective 
temperature to predict expected rates of period change for Cepheids of 
different period. A starting point is the well known period-density 
relation: 
\begin{displaymath}
P \rho^{\frac{\rm 1}{\rm 2}} = {P M^{\frac{\rm 1}{\rm 2}} \over 
(\frac{\rm 4}{\rm 3} \pi)^{\frac{\rm 1}{\rm 2}} R^{\frac{\rm 3}{\rm 2}}} 
= Q \; ,
\end{displaymath}
where {\it P} is the pulsation period, $\rho$ is the density, {\it M} is 
the stellar mass, {\it R} is the stellar radius, and {\it Q}, the pulsation 
constant, has a small period dependence \citep[e.g.,][]{kr61,fe67} that we 
assume here varies as $P^{\frac{\rm 1}{\rm 8}}$ based upon an empirical 
analysis by \citet{tb02}. Differentiation of the period-density relation, in 
conjunction with the standard equation for stellar luminosity, therefore 
leads to the following result: 
\begin{displaymath}
{\dot{P} \over P} = {{6 \over 7}{\dot{L} \over L}} {\rm -} {{24 \over 7}
{\dot{T} \over T}} \; \; .
\end{displaymath}
The desired quantity, the rate of period change $\dot{P}$, is obtained from 
tabulated differences in stellar luminosity and effective temperature as a 
function of age as a model star evolves through the instability strip.

For the present study we calculated values of $\dot{P}$ from the above 
relationship using computational stellar evolutionary models from a 
variety of available published sources, namely \citet{mm88}, \citet{al99}, 
\citet{ls01}, and \citet{cl04}. The published data were used to compute 
different parameters, depending upon the availability of the necessary 
information. \citet{al99} cite parameters for stars of different mass 
reaching the hot and cool edges of the instability strip, so their data 
yield information only about rates of period change near the center of the 
strip. In other cases, such as \citet{cl04}, there is sufficient time 
resolution in the output parameters to track changes in pulsation period 
across individual instability strip crossings. For the remaining 
sources \citep{mm88,ls01}, including \citet{cl04}, we calculated rates of 
period change for the intersection of the evolutionary tracks with the 
observationally delineated boundaries of the instability strip defined 
empirically by \citet{tu01}, which are close to those predicted by models 
of pulsation instability \citep{al99}, as well as for points lying within 
the strip boundaries. Pulsation periods were established using the 
period-radius relation \citep{tb02}. The present results differ from 
those obtained earlier \citep{tb01,tb03} in being tied to a larger variety 
of models with a greater range of metallicity, and by the inclusion of a 
weak period dependence for {\it Q} in the period-density relation.

The computed results on rates of period change are plotted in Fig. 2 for 
all of the accessible models. Different symbols denote the different 
sources. Values calculated from the models of \citet{al99} are plotted 
using filled circles, while others are plotted using open circles. Plus 
signs indicate results calculated for stars evolving through the hot and 
cool edges of the instability strip, with the rate of period change in 
general being larger on the hot edge of the instability strip, i.e., 
for more massive stars. Large symbols denote stars of solar metallicity, 
$Z = 0.02$, intermediate-sized symbols denote stars with metallicities 
of $Z = 0.01$ and $Z = 0.008$, and small symbols denote stars of very low 
metallicity, $Z = 0.001$ and $Z = 0.004$. Lines have been drawn to enclose 
those regions within which the results for different crossing modes appear 
to cluster. Sequences of points indicate models for which the time 
resolution was fine enough to calculate rate of period change over the 
entire crossing of the instability strip.

The distribution of data points in Fig. 2 suggests a variety of different 
conclusions regarding the models. First, the different models for the rapid 
first crossing of the instability strip are in very good agreement, and 
display very little variation with metallicity. The first crossing of the 
strip is a rapid transition for all stars, regardless of individual 
differences in rotation rate, etc., and that is evident from the models. 
Evidently the computational codes used for calculating the phases of shell 
hydrogen burning in stars, while perhaps differing in detail from one 
source to another, generate nearly identical results, the small variation 
in rate of period change at specific pulsation period arising from the finite 
width of the instability strip and the fact that more massive stars cross 
the strip at a greater luminosity and at a faster rate than less massive 
stars. For stars in the first crossing of the strip, high rate of period 
increase at specific pulsation period corresponds to stars on the hot edge 
of the strip, low rate of period increase to stars on the cool edge of the 
strip.

Negative period changes arise during the second crossing of the instability 
strip, which occurs during the blue loop phase of stellar evolution following 
the onset of core helium burning. The extent of the blue loop can depend upon 
a variety of factors \citep[see, for example,][]{be85,xl04}, such as 
metallicity, the treatment of core overshooting, and the distribution of CNO 
elements throughout the star. All factors affect how far a star enters the 
instability strip during core helium burning, and presumably affects how 
rapidly it evolves within the strip. Given the potentially large differences 
in initial conditions for such stars as main-sequence objects, for example, 
large variations in initial rotation rate, one might expect real stars to 
display large variations in how far they penetrate the Cepheid instability 
stripm as core helium burning objects. Somewhat unexpectedly, there are also 
very large variations among the models stars as well.

Evidently, metallicity plays only a minor role in governing the {\it rate} 
at which stars traverse the instability strip. There is as much dependence 
on the specifics of the stellar evolutionary code used. The models of 
\citet{al99}, for example, generate faster rates of period decrease than do 
other models, despite the use of common opacity tables. Models from individual 
sources are at least internally consistent in their predictions for stars of 
different masses and for stars in all portions of the second strip crossing. 
The rates of period decrease during individual strip crossings are also very 
similar to the variations predicted on the basis of mass differences, i.e, 
predicted variations in rate of period decrease at a specific pulsation 
period are generally small, except for long period Cepheids.

The third crossing of the instability strip occurs during the late stages of 
core helium burning, and gives rise to period increases, for which the 
predicted rates are depicted in the top portion of Fig. 2 along with those 
for the first crossing. Most of the comments regarding the second crossing 
of the strip apply equally to the third crossing. Again, metallicity seems to 
play a less important role in the predicted rates of period increase than 
differences in the evolutionary code. The models of \citet{al99} predict 
faster rates of period change (period increases in this case) than do other 
models, although with less consistency for stars of different mass. The rates 
of period increase during individual strip crossings are also similar to the 
variations predicted on the basis of mass differences, and predicted 
variations in the rate of period increase at a specific pulsation period are 
generally small.

A well-known problem arises for low-mass stars in the second and third 
crossings of the instability strip, since the blue loop phases of 
evolutionary models for stars of solar metallicity, $Z = 0.02$, do not enter 
the strip for $M < 4.75 M_{\sun}$ \citep[see][]{al99}. Model stars of lower 
metallicity can traverse the strip at smaller masses, but often only on the 
cool edge. By inference, most classical Cepheids of near-solar metallicity 
should have pulsation periods in excess of $\sim 3 \frac{1}{2}$ days 
\citep[e.g.,][]{tu96}, consistent with the observational sample. Nearby 
Milky Way Cepheids have abundances close to the solar values 
\citep[e.g.,][]{an02a,an02b}, and only a few have periods of less than 
$3 \frac{1}{2}$ days. Many may be overtone pulsators.

The observational picture is illustrated in Fig. 3, which presents available 
data on period changes for over 200 Cepheids, as obtained from the literature 
\citep{bp94a,bp94b,bp95,be97,tu98,bi00,be03} and ongoing research studies by 
the authors \citep[e.g.,][]{bt04,be04}. The relationships plotted in Fig. 3 
depict the regions within which the model calculations appear to cluster.

It has been pointed out previously \citep[e.g.,][]{sz83,fe84,tu98} that the 
observed rates of period change in Cepheids are generally a good match to 
predictions from stellar evolutionary models. The data of Fig. 3 provide 
further confirmation of that conclusion. Moreover, three further conclusions 
can be reached. First, once consideration is taken of the expected changes 
arising from evolution through the instability strip, the observed period 
changes in Cepheids are unlikely to contain any sizable component arising 
from another source. There are only a few exceptions to such a conclusion, 
and they are rather unusual objects like V1496 Aql \citep{be04}, which 
exhibits period changes dominated by random fluctuations in pulsation period.

Second, the observed period changes in Cepheids deviate in small but 
important ways from what is expected according to predictions based upon 
specific stellar evolutionary models. The models of \citet{al99}, for example, 
predict faster second and third crossings of the strip than those observed, 
and at much different rates. In contrast, the observed period changes in 
Cepheids are very similar for objects likely to be in the second and third 
crossings. The models of \citet{cl04} are most consistent with observations 
in that regard, but it is necessary to have a more complete mass grid of 
models constructed in the same manner to make a more detailed comparison.

Third, the range in observed rates of period change for most Cepheids is 
smaller than that resulting from a comparison of the results from different 
evolutionary models. That is somewhat surprising, given our previous 
discussion about potentially wide variations in initial conditions for 
Cepheid predecessors. Evidently real stars are similar enough in their 
internal characteristics that they evolve at fairly similar rates through 
the Cepheid instability strip.

The proportions of Cepheids in different crossing modes and in different 
period ranges in Fig. 3 are also reasonably consistent with evolutionary 
expectations. For example, stars in the first crossing of the instability 
strip during shell hydrogen burning are evolving about two orders of 
magnitude faster than stars in second and third crossings, so their relative 
numbers should be small. The two Cepheids in Fig. 3 undergoing large rates 
of period increase and falling in the predicted region for first crossers 
are Polaris ($\alpha$ UMi) and DX Gem. We assume that both are first 
crossers, as was also argued for Polaris by \citet{te05}. Moreover, the 
observed rate of period change for Polaris is now seen to be exactly what 
stellar evolutionary models predict for a star lying on the cool edge of 
the instability strip for first crossers.

The proportion of Cepheids with detectable parabolic trends in their O--C 
data also increases noticeably towards short pulsation periods, which is 
again consistent with the evolutionary expectation that the most abundant 
pulsators must be those evolving most slowly through the instability strip. 
There is a curious anomaly in the distribution of short period Cepheids, 
where essentially no variables are found to have rates of period change 
as predicted for stars in second and third crossings of the strip at 
$P \le 3.5$ days (log $P \le 0.55$). Such stars have progenitor masses of 
less than $\sim 4$ $M_{\sun}$ \citep{tu96}, where stellar evolutionary 
models for solar metallicity stars predict that the evolutionary tracks for 
core helium burning stars should no longer enter the strip. The short 
period cutoff in the observational sample is therefore consistent with 
expectations from stellar evolutionary models. But the existence of stars 
of $P \le 3.5$ days with rates of period change roughly an order of 
magnitude faster than predicted for stars in second and third crossings of 
the instability strip is not. The uncertainties in the observed rates of 
period change in the anomalous objects are generally much too small to 
resolve the anomaly by invoking systematic errors in the values of $\dot{P}$.

An additional factor that can be important for short period Cepheids is 
overtone pulsation. The Cepheids in the observational sample have all been 
assumed to be fundamental mode pulsators, and require a displacement of 
$+0.15$ in log $P$ to establish their proper locations in Fig. 3 if they 
are overtone pulsators. Yet the application of such corrections to all of 
the anomalous objects does not affect their distribution significantly; 
most still fall outside the region of $\dot{P}$-space predicted for stars 
in the second and third crossing of the instability strip. Current stellar 
evolutionary models are therefore unable to explain the existence of such 
stars, which suggests that the manner of treating the details of stellar 
evolution during blue loop stages is very important \citep[see also][]{xl04}. 
That is one area where improvements to the observational sample on Cepheid 
period changes can play an important role in testing the results from 
stellar evolutionary models.

\section{$\dot{P}$ as a Fundamental Parameter}

In Fig. 3 the dispersion in the rates of period change $\dot{P}$ observed 
in long period Cepheids ($P > 10^{\rm d}$) is smaller than what is observed 
for the calculated dispersion in that parameter among different stellar 
evolutionary models. One might expect $\dot{P}$ to correlate closely with 
location in the instability strip for Cepheids in all strip crossings, 
according to the results of Fig. 1. It is informative to examine the 
observational data more closely to determine if that is the case.

As a first step, we note that the observed rates of Cepheid period change 
plotted in Fig. 3 fall mainly within specific bands delineated by linear 
margins of slope 3.0 separated by an order of magnitude range in $\dot{P}$. 
Fig. 4 is a separate plot of the data that displays such empirically 
defined margins. All but two of the long period Cepheids with increasing 
periods fall within the lower set of margins, as do the majority of short 
period Cepheids with increasing periods. Cepheids with decreasing periods 
display a greater dispersion in $\dot{P}$ that may be intrinsic, or may be 
caused by larger uncertainties in $\dot{P}$ for the stars, particularly 
those with small rates of period change.

The anomaly for Cepheids with $P \le 3.5$ days (log $P \le 0.55$) is again 
apparent in Fig. 4. All Cepheids of shorter period display faster rates of 
period change than is typical of variables populating the lower band, and 
there are a number of stars of longer period also falling in this region of 
rapid period change. Presumably those stars represent Cepheids in fourth and 
fifth crossings of the instability strip, with faster associated rates of 
period change. Multiple crossings of the instability strip appear to be 
possible for stars in late core helium burning stages, depending upon the 
CNO abundances in the hydrogen burning shells of such stars \citep{xl04}.

The finite range in stellar surface temperature for stars populating the 
instability strip at constant pulsation period implies distinct differences 
in pulsation efficiency that should coincide with marked differences in 
pulsation amplitude for Cepheids of similar period. On the hot edge of the 
strip the ionization zone is just beginning to reach depths where the 
piston mechanism for pulsation becomes efficient, so light amplitudes should 
be small but increasing with decreasing surface temperatures. On the cool 
edge the lower surface temperatures are associated with increased convective 
energy transport in the star's outer layers \citep{de80}, so pulsation 
amplitudes should also be small. 

The first study of Cepheid amplitudes as a function of position in the strip 
by \citet{kr63} was consistent with that picture, although small amplitude 
Cepheids were found only on the hot edge of the strip. All subsequent 
amplitude maps of the instability strip by \citet{ho67}, \citet{st71}, 
\citet{pg74}, \citet{pl78}, \citet{tu01}, and \citet{sa04} have produced 
similar results, namely a sharp rise to maximum amplitude on the hot edge of 
the strip followed by a more gradual decline towards the cool edge.

Cepheid amplitudes display a period dependence as well as a dependence upon 
location within the strip, a natural consequence of an effect tied to 
surface gravity as well as pulsation efficiency. In order to eliminate that 
factor in characterizing Cepheid period changes, we have normalized the 
resulting values of blue light amplitude and $\dot{P}$ as follows: (i) blue 
amplitudes $\Delta B$ were standardized through the ratio $\Delta B$/$\Delta 
B$(max), where $\Delta B$(max) is the maximum value of $\Delta B$ for the 
star's pulsation period, and (ii) $\dot{P}$ was adjusted to the equivalent 
value for a Cepheid with a pulsation period of $10^{\rm d}$ using the 
empirically-obtained slope plotted in Fig. 4.

Fig. 5 plots such data for Cepheids with $12^{\rm d} \leq P \leq 40^{\rm d}$ 
and increasing pulsation periods ($P \simeq 20^{\rm d}$). The upper part of 
the diagram plots the individual data, while the middle part of the diagram 
plots running means for the data. The lower part of the diagram is an 
alternate interpretation of the same data, as described below. Similar plots 
are given in Fig. 6 for Cepheids with $4^{\rm d} \leq P \leq 8^{\rm d}$ and 
increasing pulsation periods ($P \simeq 6^{\rm d}$), and in Fig. 7 for 
Cepheids with $4^{\rm d} \leq P \leq 8^{\rm d}$ and decreasing pulsation 
periods ($P \simeq 6^{\rm d}$). Recall that large values of $\dot{P}$ should 
correspond to the hot side of the instability strip, and small values to 
the cool side.

The data for $20^{\rm d}$ Cepheids (top portion of Fig. 5) display a 
tendency for large amplitude Cepheids to have rates of period increase 
typical of stars lying near the center of the instability strip, with smaller 
amplitude Cepheids falling towards the hot and cool edges (larger and smaller 
values of $\dot{P}$, respectively). The trend is more obvious when one 
plots running five-point means of the same data, as in the middle section 
of Fig. 5. There are two long period Cepheids with anomalously large values 
of $\dot{P}$, SZ Cas and AQ Pup, which are conceivably fifth crossers. If 
they are omitted from the running means and averages over smaller samples are 
included at the extremes of $\dot{P}$, one obtains the results in the lower 
portion of Fig. 5, which are typical of independent cross-sectional amplitude 
maps of the instability strip. The scatter in $\dot{P}$ values evident in the 
top portion of Fig. 5 is intrinsic to the stars, and is not the result of 
large uncertainties in the calculated values. Presumably there are intrinsic 
physical differences from one Cepheid to another that account for the scatter, 
as noted earlier. Differences in initial rotation velocity for the progenitor 
main-sequence stars might be the sole factor, given that they would generate 
sufficiently large variations in the abundances of the CNO elements throughout 
the star to affect the extent of the blue loop stages \citep{xl04}.

The data for $6^{\rm d}$ Cepheids with period increases (top portion of 
Fig. 6) are more complicated. It appears that the sample consists of two 
overlapping groups of objects, a feature that also appears in the running 
five-point means displayed in the middle section of Fig. 6. We assume that 
each group consists of Cepheids displaying an order of magnitude (factor 
of 10) variation in $\dot{P}$ values from the hot to cool edges of the 
instability strip, as displayed by the long period Cepheids in Fig. 5, and 
use the results presented in the lower portion of Fig. 5 as a template for 
the likely variations in relative amplitude with $\dot{P}$ for short period 
Cepheids. When the two groups in Fig. 6 are separated in such fashion and 
averages over smaller samples are included at extreme values of $\dot{P}$ 
for each group, one obtains the results in the lower portion of Fig. 6. The 
simplest explanation for the existence of two groups among the short period 
Cepheids is the existence of higher strip crossings among the stars, namely 
fifth crossings for Cepheids undergoing period increases.

Similar results apply to the data for $6^{\rm d}$ Cepheids with period 
decreases (Fig. 7), when analyzed in similar fashion, despite the smaller 
sample size. The top portion of Fig. 7 displays excessive scatter, much 
like that in the top portion of Fig. 6, with only marginal improvement 
through running five-point means (middle portion of Fig. 7). Restricting 
the data sets as above produces the results depicted in the lower portion 
of Fig. 7, which suggests an overlap between Cepheids in second and fourth 
crossings of the instability strip. As noted for long-period Cepheids, 
there is an intrinsic scatter in $\dot{P}$ values that is not the result 
of large uncertainties in the calculated values. Such scatter makes it 
difficult to use rate of period change for individual Cepheids to identify 
their exact location within the instability strip, although approximate 
placements are possible in most cases.

The conclusions reached here, while admittedly speculative, provide a 
reasonable explanation for the characteristic behavior of pulsation 
amplitude and rate of period change for individual Cepheids at specific 
pulsation period. As noted earlier, there may be further complications 
arising from the presence of overtone Cepheids in the sample, but their 
numbers should be relatively small in the present case, except at short 
periods, and they would not alter the observed distribution of data 
points significantly. A more comprehensive study including recognized 
overtone Cepheids should be possible once O--C studies have been 
completed for such variables.

\section{Discussion}

Our intent here is to demonstrate that rate of period change for a Cepheid 
is a useful parameter that permits one to characterize the variable in 
terms of specific evolutionary state. Information on $\dot{P}$ for a Cepheid, 
in conjunction with its known pulsation period and light amplitude, can be 
used to identify the strip crossing mode for the object as well as its 
likely location within the strip, the latter independent of its observed 
color and reddening. The parameter $\dot{P}$ may even be useful for 
establishing if a Cepheid is a fundamental mode pulsator or an overtone 
pulsator, although we leave that as a future exercise.

If we interpret the results of Figs. 5--7 as a generic indicator of how 
pulsation amplitude varies across the instability strip, then the width of 
the strip in $\dot{P}$ at constant period, which amounts to $\sim 1.2$ in 
log $\dot{P}$, must encompass a range of $\sim 16$ in $\dot{P}$. Of that, 
an intrinsic dispersion in log $\dot{P}$ values amounting to perhaps 
0.4--0.5, a factor of $\sim 3$, presumably arises from actual internal 
differences in the Cepheids resulting from different histories for their 
progenitor stars. Specific stellar evolutionary models presented in 
Fig. 2 predict a smaller variation in log $\dot{P}$ than what is observed, 
which may reflect the simplicity of the models. In that regard, observed 
rates of period change in Cepheids can play an important role as a check 
on how closely stellar evolutionary models match real stars. Until now 
Cepheid period changes have not been used for that purpose.

\acknowledgments

The present study was supported by research funding awarded through the 
Natural Sciences and Engineering Research Council of Canada, the Russian 
Foundation of Basic Research through the Federal Program "Astronomy" of 
the Russian Federation, and the Small Research Grants Program of the 
American Astronomical Society, in part by funding from the Cecelia Payne 
and Sergei Gaposchkin Memorial Fund.

\clearpage
\begin{figure}
\plotone{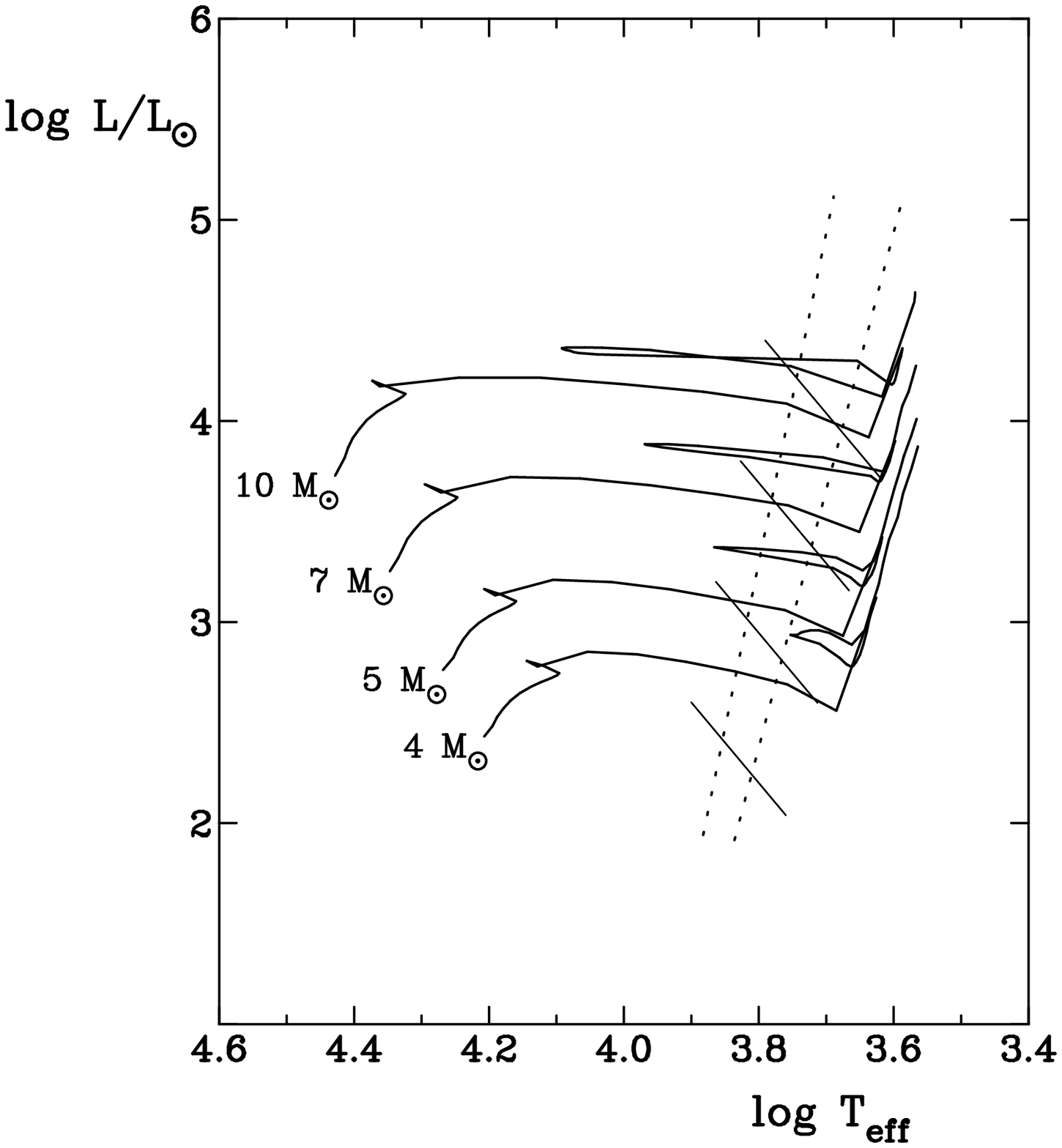}
\caption{The theoretical HR diagram illustrating post-main-sequence 
evolutionary tracks for stars of 4, 5, 7, and 10 $M_{\sun}$ \citep{ls01}. 
Included is the observational location of the Cepheid instability strip 
\citep[dotted lines, from][]{tu01} and lines of constant stellar radius. 
\label{fig1}}
\end{figure}

\clearpage
\begin{figure}
\plotone{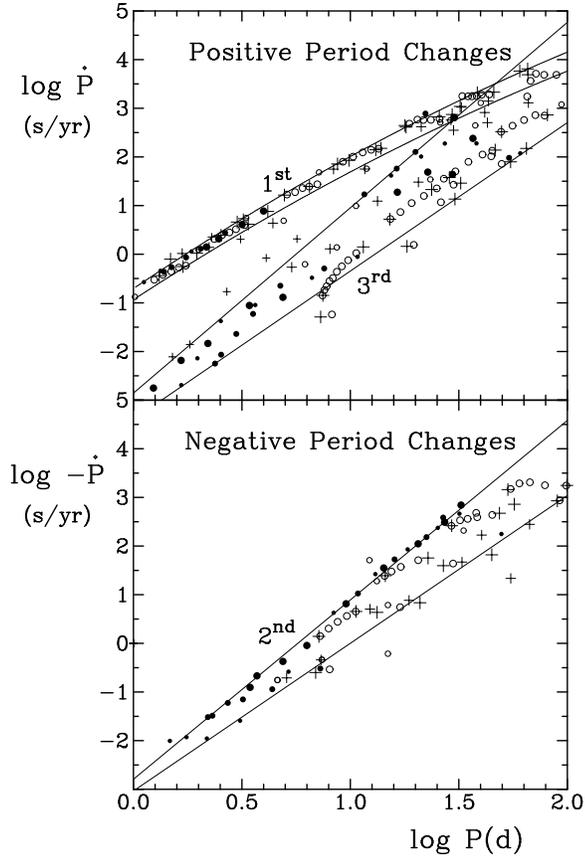}
\caption{Predicted rates of period change for stars crossing the Cepheid 
instability strip as tied to published stellar evolutionary models. The 
meaning of the different symbols is explained in the text. Lines denote 
regions within which the predictions from different stellar evolutionary 
models appear to cluster. The different crossings of the instability 
strip are identified. \label{fig2}}
\end{figure}

\clearpage
\begin{figure}
\plotone{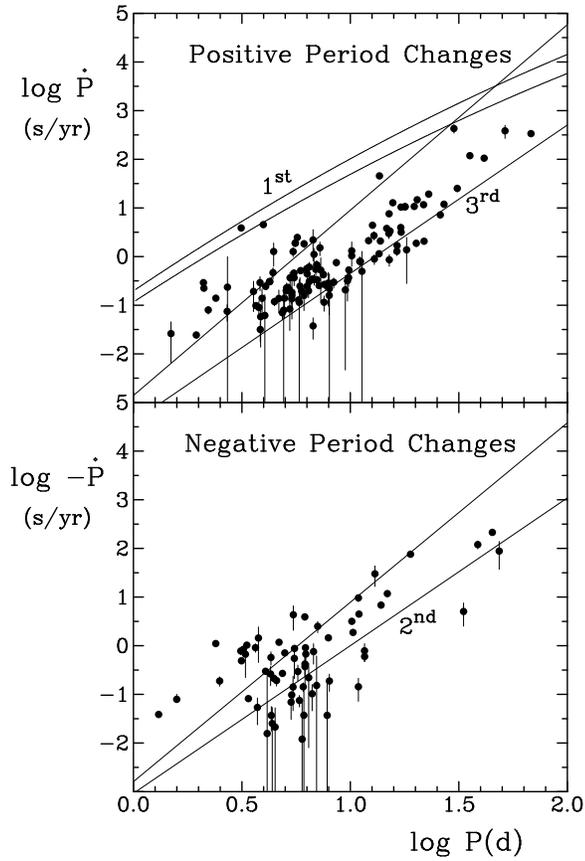}
\caption{Observed rates of period change, along with their calculated 
uncertainties, for well-studied Cepheids possessing many years of O--C 
data. Dotted lines are the relations depicted in Fig. 3, and the different 
strip crossings are identified. \label{fig3}}
\end{figure}

\clearpage
\begin{figure}
\plotone{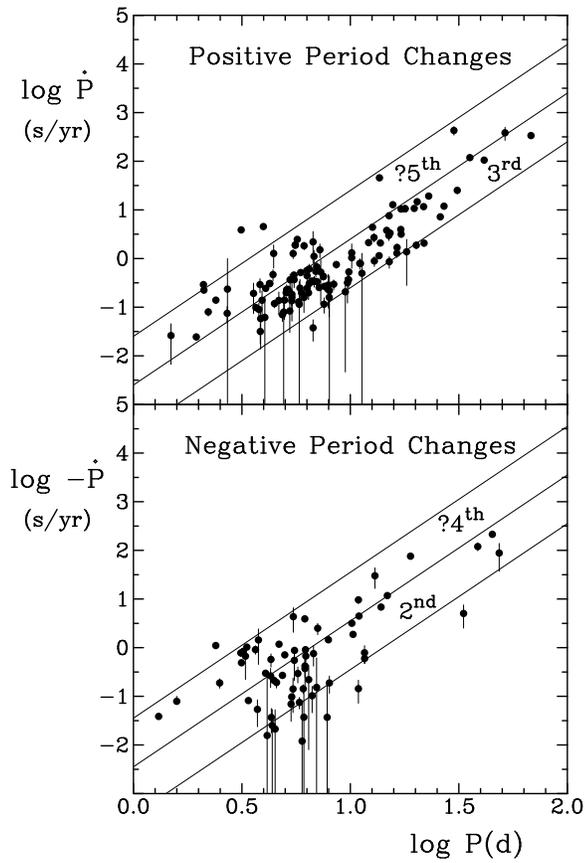}
\caption{The data of Fig. 3 plotted along with suggested empirical 
delineations of the regions corresponding to Cepheids in different crossings 
of the instability strip, as identified. \label{fig4}}
\end{figure}

\clearpage
\begin{figure}
\plotone{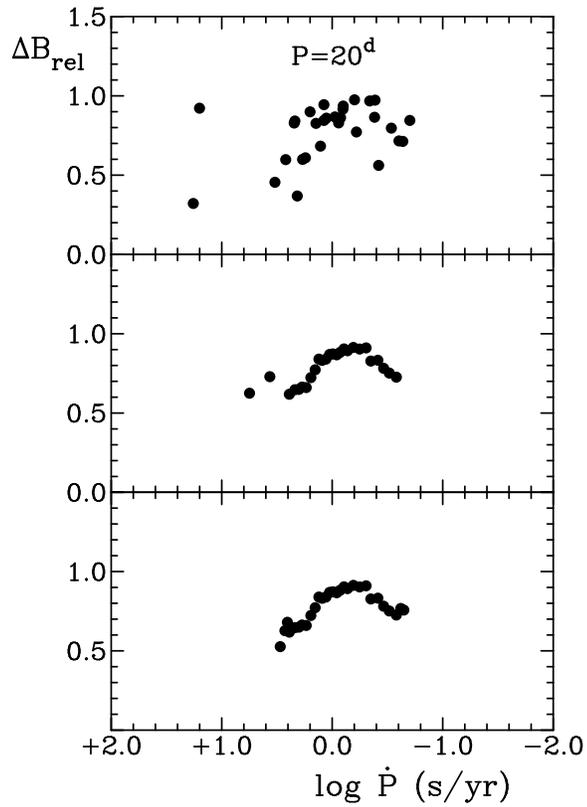}
\caption{Normalized blue amplitudes of Cepheids with $12^{\rm d} \leq P \leq 
40^{\rm d}$ and increasing pulsation periods as a function of normalized rate 
of period change (upper section). The middle section displays running 
five-point means for the data, and the lower section adjusted and extended 
means of the data. \label{fig5}}
\end{figure}

\clearpage
\begin{figure}
\plotone{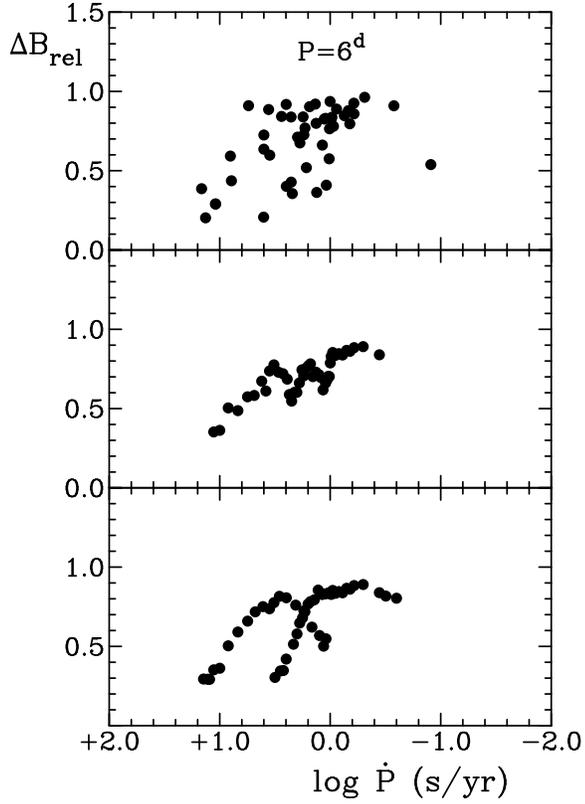}
\caption{Normalized blue amplitudes of Cepheids with $4^{\rm d} \leq P \leq 
8^{\rm d}$ and increasing pulsation periods as a function of normalized rate 
of period change (upper section). The middle section displays running 
five-point means for the data, and the lower section adjusted and extended 
means of the data after judicious separation of the overlapping samples using 
the results of Fig. 5 as a template. 
\label{fig6}}
\end{figure}

\clearpage
\begin{figure}
\plotone{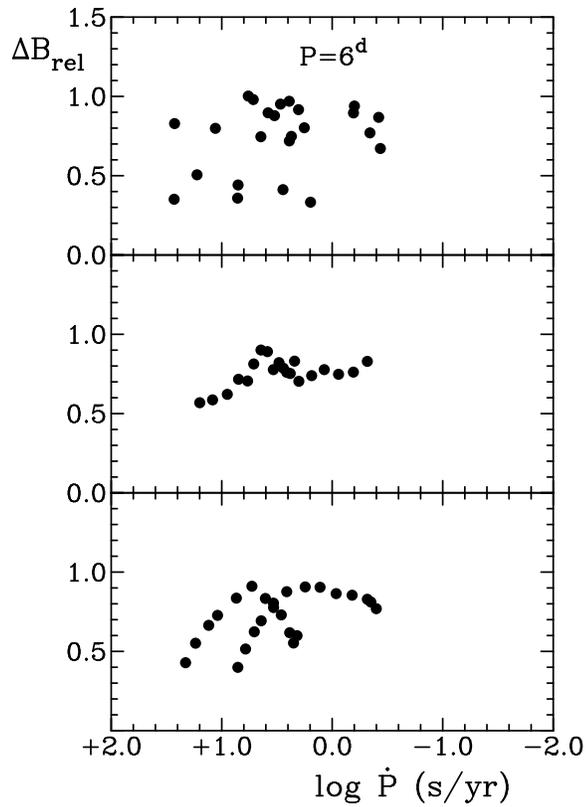}
\caption{Normalized blue amplitudes of Cepheids with $4^{\rm d} \leq P \leq 
8^{\rm d}$ and decreasing pulsation periods as a function of normalized rate 
of period change (upper section). The middle section displays running 
five-point means for the data, and the lower section adjusted and extended 
means of the data after judicious separation of the overlapping samples as 
in Fig. 6. 
\label{fig7}}
\end{figure}

\end{document}